# Femtosecond THz time domain spectroscopy at 36 kHz scan rate using an acousto-optic delay

B. Urbanek[1], M. Möller[1], M. Eisele[1], S. Baierl[1], D. Kaplan[2], C. Lange[1*], and R. Huber[1]

[1]*Department of Physics, University of Regensburg, 93053 Regensburg, Germany*

[2]*Fastlite, 1900 route des crêtes, 06560 Valbonne, France*

[*]*christoph.lange@physik.uni-regensburg.de*

**We present a rapid-scan, time-domain terahertz spectrometer employing femtosecond Er:fiber technology and an acousto-optic delay with attosecond precision, enabling scanning of terahertz transients over a 12.4-ps time window at a waveform refresh rate of 36 kHz, and a signal-to-noise ratio of $1.7 \times 10^5/\sqrt{\text{Hz}}$. Our approach enables real-time monitoring of dynamic THz processes at unprecedented speeds, which we demonstrate through rapid 2D thickness mapping of a spinning teflon disc at a precision of $10$ nm/$\sqrt{\text{Hz}}$. The compact, all-optical design ensures alignment-free operation even in harsh environments.**

Terahertz photonics has opened up a vast variety of applications in disciplines as diverse as physics, chemistry, materials sciences, medicine, and biology [1-4]. In THz time-domain spectroscopy, phase-locked electromagnetic pulses have been routinely generated, and amplitude and phase of their oscillating carrier wave have been detected by electro-optic gating. The necessary temporal resolution has typically been implemented via movable mechanical actuators controlling the delay between ultrashort pulses. Few-cycle THz transients have probed the dynamics of elementary excitations in solids [1, 3] such as plasmas [5, 6], excitons [7], phonons [8], magnons [9], or spin currents [10]. Whereas the repeatable nature of these phenomena allows for stroboscopic sampling of the THz field in a laboratory environment, many industrial or biological applications require real-time acquisition of singular events at kHz scan rates, facilitating rapid DNA analysis [11], spectroscopy of biomolecules [12, 13], environmental gas sensing [2, 14], drugs and explosives testing [2, 14, 15] or imaging applications [16]. Hence, numerous strategies towards replacing mechanical delays by faster, more precise and more robust concepts have evolved. First approaches included loudspeaker diaphragms [17] or rotating mirrors [18, 19], which increased acquisition speeds to a few tens of Hz. All-optical schemes such as asynchronous optical sampling (ASOPS) [20], electrically controlled optical sampling (ECOPS) [21-23], or optical sampling through cavity tuning (OSCAT) [24] have greatly advanced the field and enabled scan rates of up to 8 kHz [22]. Specific challenges of these techniques include jitter and dead times when using two cavity-stabilized lasers (ASOPS), dispersion of long fiber branches (OSCAT), or complex intra-cavity elements in the laser source (ECOPS).

More recently, an acousto-optic programmable dispersive filter (AOPDF) was used to implement a novel femtosecond optical delay reaching scan rates of 36 kHz with attosecond precision [25]. The scheme is based on acousto-optic interaction of a mechanical wave with co-propagating laser pulses in a birefringent crystal [26]. The delay generates gapless pulse trains consisting of ~$10^3$ pulses with a delay increment of a few femtoseconds imposed on successive pulses. Its all-optical design, low dispersion, and attosecond precision render it ideal for phase-resolved spectroscopy [27] and hold the potential to overcome most limitations of existing delay concepts for THz spectroscopy, enabling rapid-scan time-domain spectroscopy from below 1 THz up to the optical domain.

Here, we exploit this delay to implement a THz time-domain spectrometer based on Er:fiber laser technology, achieving a THz waveform update rate of 36 kHz and a scan range of 12.4 ps, at a resolution of 11.3 fs. Electro-optic sampling of the THz field aided by field-programmable gate-array (FPGA) electronics synchronized to the laser repetition rate allows for recording and processing one THz transient every 28 µs at a signal-to-noise-ratio (SNR) of 27, corresponding to a normalized SNR of $1.7 \times 10^5/\sqrt{Hz}$. As a first experiment, we perform rapid, two-dimensional mapping of the topology of a rotating teflon disc by THz transmission measurements with a precision of 10 nm/$\sqrt{Hz}$.

Our Er:fiber oscillator (Figure 1a, OSC) generates 92-fs pulses at a center wavelength of 1.55 µm and a repetition rate of $f_{rep}$ = 40 MHz [28]. The pulses are amplified to an energy of 8 nJ (AMP 1) and represent the reference pulse train (1). A fraction of 1 % of the laser output is split off to branch (2) and fed through the 25-mm-long $TeO_2$ crystal of the AOPDF (model: Fastlite Dazzler) with its polarization aligned along the ordinary axis of the crystal. An acoustic wavepacket is launched synchronously to the laser at a rate of $f_a$ = 36 kHz, resulting in $N = f_{rep}/f_a \approx 1100$ laser shots passing through the crystal within a single AOPDF period. Each laser pulse encounters the wavepacket at an increasing distance along the crystal axis (Fig. 1b), where it is Bragg-scattered to the extraordinary axis and continues to travel at a slower group velocity. After passing the crystal once, each laser pulse has accumulated a delay $\Delta t_i = (N-i) \times \Delta t_0$ relative to its counterpart in branch (1), where $i$ is the index of the laser pulse after the start of the AOPDF period, and $\Delta t_0$ is the fundamental delay increment which develops between two successive laser pulses [25, 26]. The acoustic waveform sent to the AOPDF is shaped to compensate for the spectral dispersion of the $TeO_2$ crystal. This allows us to double the delay increment and, thus, the scan range by reflecting the beam back through the crystal under a small angle relative to the first pass, and obtain pulse trains with a successive delay of 11.3 fs, covering a scan range of 12.4 ps. Phase resolved autocorrelation measurements taken before (Fig. 1c, red curve) and after passing the crystal (Fig 1c, dashed gray envelope) indeed confirm that the effects of dispersion are negligible, as the average pulse duration increases only slightly from 92 to 100 fs despite the considerable bandwith (FWHM) of our laser of $\approx 80$ nm (Fig 1d). We note that the high scattering efficiency of the AOPDF of $\approx 35\%$ suggests that in the future, even three- or four-pass configurations may be used to further extend the delay. After passing the AOPDF, the pulses seed the second amplifier (Fig. 1a, AMP 2), which boosts the pulse energy to 8 nJ. The amplified pulses generate phase-stable single-cycle THz transients through optical rectification in a nonlinear, organic phenolic polyene crystal (OH1) [29]. From this point on, the THz beam is housed in a purge box with a dry atmosphere to avoid THz absorption by water vapour. Using off-axis parabolic mirrors, the THz field is collimated, focused to an intermediate spot for sample analysis, and recollimated. Finally, the transient is electro-optically gated in a ZnTe crystal by the reference pulses. For optimized phase-matching in the detector crystal, the gate pulses are frequency-doubled to a wavelength of 775 nm in a $LiNbO_3$ crystal. Polarization optics are employed for balanced detection using two photodiodes, which are read out shot by shot using high-precision 16-bit A/D converters controlled by FPGA electronics locked to the laser repetition rate and triggered synchronously to the acoustic wave. Within one period of the AOPDF of $t_{AO} \approx 28$ µs, a complete THz waveform is recorded, buffered, and subsequently transferred to a computer for statistical analysis while acquisition continues seamlessly.

The electric field $\mathcal{E}$ of our single-cycle THz waveform is plotted in Fig. 2 for different acquisition times ranging from $\tau = 28$ µs (top) to 1 s (bottom), which is equivalent to averaging over 1 up to $3.6 \times 10^4$ sweeps of the AOPDF delay. The spectrum of our waveform retrieved within $\tau = 1$ s (inset, black curve) expands from 0.1 to 4 THz with a broad plateau between 0.7 and 2.6 THz suitable for high-contrast absorption measurements. Even at an acquisition rate of 36 kHz, the spectral amplitude (shaded gray area) lies well above the noise floor over a range from 0.8 to 2.8 THz, rendering our setup capable of THz spectroscopy on the microsecond time scale. Restricting the bandwidth of our spectrometer to below 5 THz (red curve), we achieve a signal-to-noise ratio (SNR) of $\approx 27$ already for a single sweep of the AOPDF ($\tau = 28$ µs), corresponding to a SNR of $1.7 \times 10^5/\sqrt{Hz}$ when normalized to an individual electro-optic delay time. When we switch to the full bandwidth suitable for multi-THz spectroscopy up to 50 THz, we still obtain a SNR of $6.3 \times 10^4/\sqrt{Hz}$. Apart from these real-time capabilities, our flexible data acquisition allows for various measurement techniques which go beyond simple averaging methods. In box-car or multi-dimensional pump-probe scenarios, data analysis may include simultaneous demodulation at multiple frequencies associated with different triggers. Monitoring singular, non-repetitive events can be performed at the full waveform update rate of 36 kHz while the averaging time may be adapted during post-processing, depending on the real-time dynamics of the THz signal.

Figures 3a and b illustrate the stability of our setup. The two-dimensional color plots represent the THz field as a function of electro-optic delay time and real time over two 20-ms segments at the start and the end, respectively, of a 90-s recording interval. We determine the time zero of every THz transient directly from the electro-optic signal by performing maximum-likelyhood-analysis with an averaged reference transient, thus exploiting the entire time-domain data for the fit. In Fig. 3c, we plot the histogram of the deviation of the time zero from its mean value, $\Delta t_{d0}$. The distribution is characterized by a standard deviation of 2.7 fs, currently constrained only by the THz SNR. For stronger THz fields, we expect this figure to ultimately improve towards the 15-as precision limit of the acousto-optic delay, with a long-term absolute precision below 1 fs even without active stabilization [25]. At the current level of precision, free-space ranging applications in air would already yield a resolution as good as 2.1 nm/$\sqrt{Hz}$. Likewise, real-time monitoring of the thickness or composition of materials sufficiently transparent to THz radiation such as polymers or cellulose would be possible already during manufacturing, at high throughput and reliability. The high SNR suggests that transfer-matrix analysis of mm-thick multi-layer structures [30] may be performed even for material stacks with sub-wavelength spacing and a low contrast of the refractive index.

To demonstrate the practicality of our approach for real-time thickness profiling we apply the above maximum-likelihood procedure to measure the thickness of the outer perimeter of a teflon disc with a diameter of 100 mm. The disc exhibits an angular thickness variation over a total of 180 equally spaced segments, modeled after the silhouette of the medieval city of Regensburg, whereby the thickness is constant within each segment (Fig. 3d). The perimeter of the disc is placed in the THz focal spot while the disc is spun at 2670 rpm, rapidly modulating the optical path length for the THz transient according to the refractive index contrast between teflon and air, $n_{teflon} - n_{air} = 0.42$. In Fig. 3e, we plot the electric field analogously to Fig. 3a for a complete revolution of the disc, recorded in only 23 ms. The temporal offsets of the THz transients trace the silhouette with high definition, rendering even steep thickness gradients. In the inset, a magnified view of the data marked by the gray rectangle is plotted. Here, the step-shaped structure of the roof of the cathedral, in which the thickness varies notably between adjacent angular segments, is clearly resolved with 5 field transients per segment. Between two steps, the travel time of the THz transient is defined by the spatial average of the thickness, and its amplitude is reduced due to diffraction, leading to a slightly smeared-out signature. Taking into account the refractive index, a statistical analysis of the precision of our method yields a standard deviation of 1.9 µm for a single transient, equivalent to $10\,\mathrm{nm}/\sqrt{\mathrm{Hz}}$.

Such high precision suggests that even two-dimensional thickness mapping should be feasible in a short time frame. To this end, we swap the OH1 emitter with a 500-µm-thick, (110)-cut GaAs crystal, which generates considerably weaker THz transients but allows for more resilient operation under rough environmental conditions at a fraction of the cost of its organic counterpart, and perform a topology scan of a second teflon disc with a two-dimensional thickness variation. The disc is partitioned into more than $10^4$ square elements of a size of $1 \times 1\,\mathrm{mm}^2$, each, modeling the topology of the greater Tokyo area including Mt. Fuji. We mount the disc on a linear stage to move it uniformly through the THz focus while spinning it at 2670 rpm. After maximum-likelihood analysis, a cylindrical mapping function is used to reconstruct the 3D-height profile from the linear THz data stream. An angular marker on the disc allows for orientation within the data and correcting for fluctutions of the angular velocity. In the top part of Fig. 3f, the measured topology is rendered under an oblique angle, accurately revealing the bay area and the symmetrical cone of the volcano. The template is shown for reference below. In the total acquisition time of $\approx 20$ s, more than $7 \times 10^5$ height samples were determined.

The above examples show that THz rapid scan technology based on acousto-optic delays has the potential to revolutionize imaging, ranging, and quality control in industrial applications. Yet, its potential extends much further, encompassing biological applications, allowing real-time monitoring of non-repeatable processes on the microsecond scale, or chemistry, where the frequency domain may be exploited to monitor reactions at unprecedented rates. In multi-dimensional THz spectroscopy, our scheme enables collapsing the scan time for one or more axes by orders of magnitude, reducing total acquisition times for complex experiments from hours to minutes. Especially with regard to multi-THz spectroscopy, the attosecond precision and complete lack of hysteresis of acousto-optic delays represent a key advantage over conventional mechanical actuators.

In conclusion, we have demonstrated a rapid THz time domain spectrometer based on an acousto-optical programmable dispersive filter, capable of gap-free acquisition of THz transients at an unprecedented waveform refresh rate of 36 kHz. Recent developments in AOPDF technology aim to increase the repetition rate above 100 kHz in order to match current short pulse amplifier performance. One can thus foresee even higher refresh rates in the future. The delay covers a 12.4-ps window with a temporal resolution of 11.3 fs and can be easily adapted for longer scan ranges or resolutions by crystal geometry, programming of the AOPDF, or by using multi-pass configurations. With a signal-to-noise ratio of $1.7 \times 10^5/\sqrt{\text{Hz}}$, our delay allows for investigating THz dynamics on a microsecond time scale, as we have demonstrated through thickness measurements of rapidly spinning teflon discs with a precision of $10\,\text{nm}/\sqrt{\text{Hz}}$. The extremely low dispersion and attosecond precision of the AOPDF make our concept suitable for multi-THz spectroscopy up into the near-visible regime.

The authors acknowledge support by the European Research Council through grant no. 305003 (QUANTUMsubCYCLE), and technical support by Toptica Photonics and Rainbow Photonics.

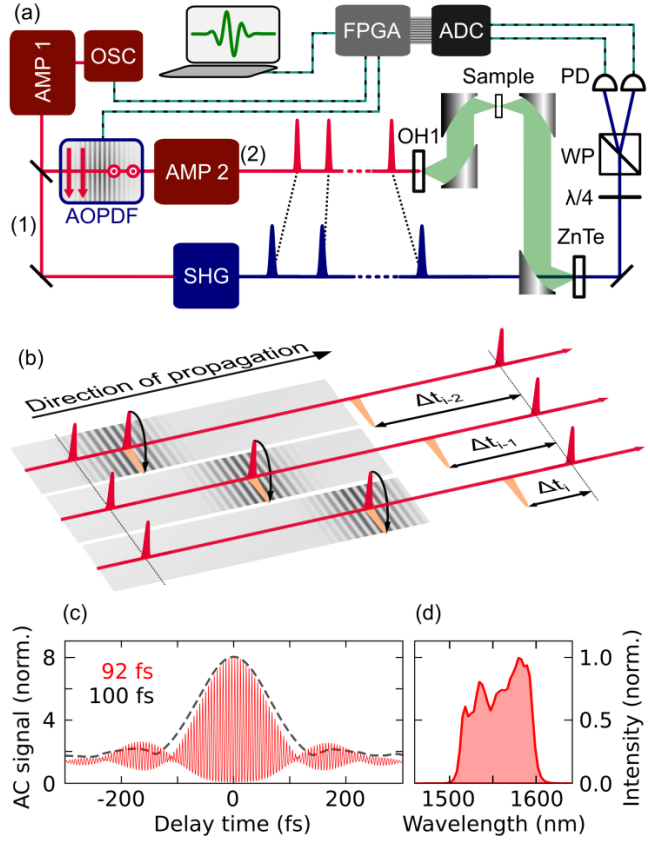

Fig. 1: (a) Schematic setup. A femtosecond 40-MHz Er:fiber oscillator and amplifier (OSC, AMP 1) provide the pulse train for electro-optic sampling in a ZnTe crystal (branch 1) and the train of pump pulses, which are delayed through the AOPDF (branch 2) and amplified in a second stage (AMP 2) for THz generation in an organic crystal (OH1). λ/4: quarter waveplate, WP: wollaston prism, PD: photodiodes, for electro-optic detection. FPGA, ADC: Acquisition electronics for seamless real-time capturing. SHG: $LiNbO_3$ crystal for frequency doubling. (b) Sketch of the AOPDF crystal (gray) and the acoustic wave (gray-shaded modulation pattern), which Bragg-scatters laser pulses from the ordinary crystal axis (red) to the extraordinary axis (orange), resulting in an accumulative relative delay (see text). (c) Interferometric autocorrelation of laser pulses before (red curve) and after transmission through the AOPDF (dashed gray envelope). (d) Spectrum of fiber laser (FWHM = 80 nm).

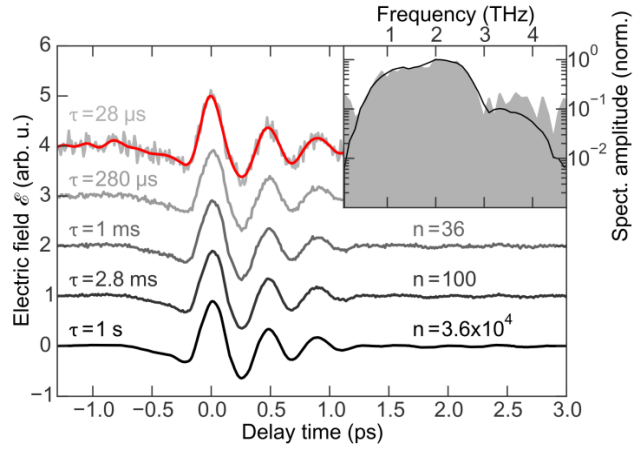

Fig. 2: THz transients obtained by averaging over an increasing number of AOPDF periods, corresponding to acquisition times of $\tau$ = 28 μs to 1 s (gray and black curves, top to bottom). Red curve: THz transient for $\tau$ = 28 μs obtained by limiting the bandwidth to below 5 THz. The signal-to-noise ratio of this single-scan transient reaches a value of 27. Inset: Frequency spectrum for $\tau$ = 28 μs (shaded gray area) and for $\tau$ = 1 s (black curve).

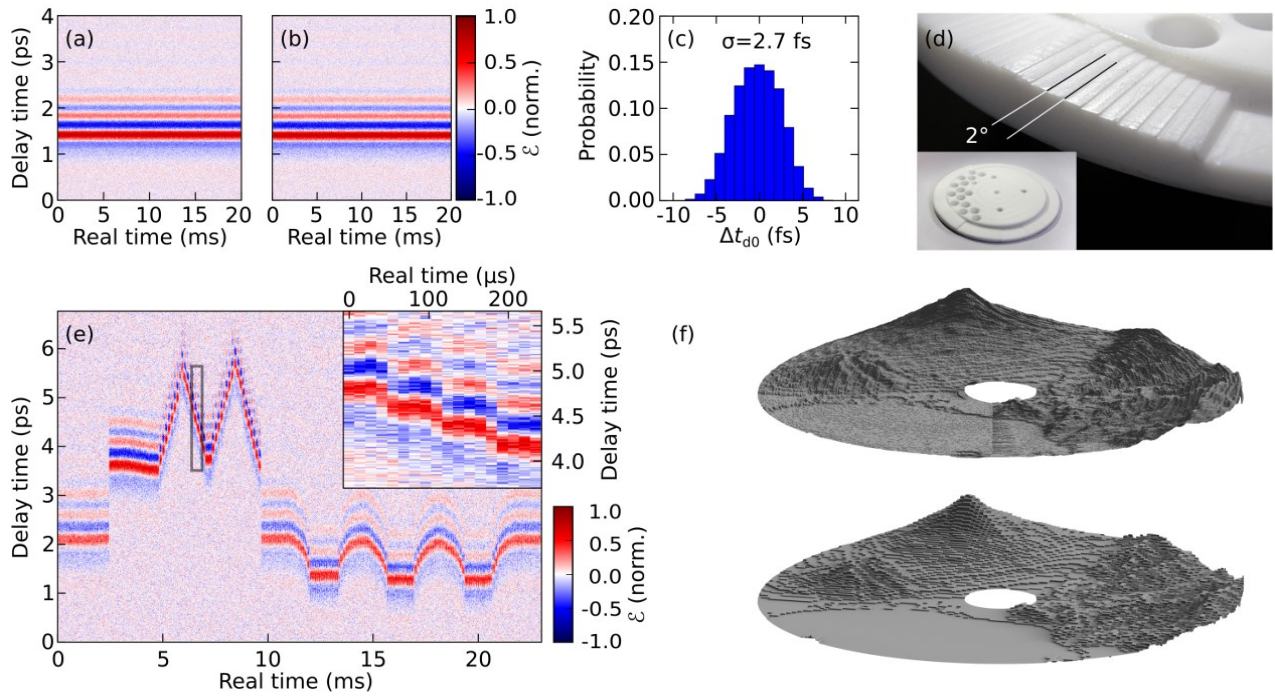

Fig. 3: (a) Color plot of THz transients recorded at a scan rate of 36 kHz over a 20 ms interaval. (b) Plot of transients equivalent to (a), recorded 90 s afterwards. (c) Histogram showing the distribution of time zeroes for transients in (a) as determined through maximum-likelihood-analysis. (d) Close-up photograph of teflon disc with thickness variation over angular segments spaced by 2°, and entire disc (inset). (e) 2D-map of THz transients during a single revolution of the disc, tracing the silhouette of the cathedral and ancient stone bridge of Regensburg. Inset: magnified view of the data outlined by the gray box. (f) Reconstructed height profile of a second teflon disc modeling the topography around Mt. Fuji (top), and the manufacturing template (bottom).